# ROOM TEMPERATURE NANOSTRUCTURED GRAPHENE TRANSISTOR WITH HIGH ON/OFF RATIO

Mircea Dragoman[1*], Adrian Dinescu[1], and Daniela Dragoman[2,3]

[1]National Institute for Research and Development in Microtechnology (IMT), P.O. Box 38-160, 023573 Bucharest, Romania,

[2]Univ. Bucharest, Physics Faculty, P.O. Box MG-11, 077125 Bucharest, Romania

[3]Academy of Romanian Scientists, Splaiul Independentei 54, 050094 Bucharest, Romania

**Abstract**

We report the batch fabrication of graphene field-effect-transistors (GFETs) with nanoperforated graphene as channel. The transistors were cut and encapsulated. The encapsulated GFETs display saturation regions that can be tuned by modifying the top gate voltage, and have on/off ratios of up to $10^8$ at room temperature. In addition, the nanoperforated GFETs display orders of magnitude higher photoresponses than any room-temperature graphene detector configurations that do not involve heterostructures with bandgap materials.

_________________________________________________________________

*Corresponding author: mircea.dragoman@imt.ro

## 1. Introduction

The absence of bandgap in graphene leads to serious detrimental effects on graphene field-effect-transistors (GFETs). In particular, there are no saturation regions and the graphene channel cannot be switched off, resulting in a poor ratio between on and off drain currents. In addition, very often, the drain conductance is larger than the transconductance, so that the GFET is not amplifying.

There are many methods to open a bandgap in graphene, but they generally destroy the electrical properties of the material, especially the mobility, which makes graphene so desirable for nanoelectronics. When graphene is not processed at all, the on/off ratio is around 5, but this parameter increases with an order of magnitude, reaching 50, when GFET is nanopatterned with a two-dimensional array of nanoperforated holes with an average diameter of 100 nm and a neck average of 10 nm [1]. An on/off ratio of $10^3$ is obtained in GFET if a nanoconstriction is made in graphene [2], it increases up to $2 \times 10^3$ in bilayer graphene GFETs [3], and reaches $10^4$-$10^5$ in suspended graphene nanoribbons [4].

However, despite the dramatic improvement of the on/off ratio, none of these GFETs is suitable for batch fabrication at wafer level, and the drain current is very weak, in the range of 1-5 μA. The reason is that in nanopatterns with neck sizes of 5-10 nm the pattern uniformity cannot be maintained on large areas and thus the reproducibility is lost. So, batch production at wafer scale is hampered by lithographic methods that try to obtain too small features. Moreover, the saturation regions are still elusive in Refs. [1-4], although a clear tendency towards the switching off of the transistors is clearly evidenced.

Graphene nanopatterning is an emerging research topic, with many applications in nanoelectronics, gas sensors, biosensors, and plasmonics (see the recent review [5]). Nanopatterning inevitably degrades the physical properties of graphene by introducing edge disorder of the nanosize holes, squares, or hexagons, which constitute the nanopatterned array. However, very high mobility and ballistic transport at room temperature with a carrier mean free path of 400 nm can be still preserved if the perforated graphene is sandwiched between hexagonal boron nitride layers [6]. There are several methods to create nanopatterns of different shapes, such as holographic lithography and template-like methods, which include block-copolymer lithography, self-assembled nanosphere masks, and aluminium oxides etching masks. A review of these methods, which can create nanopatterns with neck sizes of few nanometers, and an extensive bibliographical list are provided in [7].

The aim of this paper is to demonstrate that batch production at the wafer scale of nanopatterned GFETs is possible if e-beam lithography is used to produce in few seconds a nanoperforated graphene channel with nanopatterns having neck sizes of several tens of nanometers. The e-beam fabrication of nanoperforated holes in graphene, with a diameter of 100 nm and a period between holes of 100 nm, was used recently in the context of localization states and band bending evidence in graphene antidot superlattices [8]. All nanopatterned GFETs reported in this paper have milliamperes drain current levels, which are orders of magnitude greater than in other nanopatterned transistors, show saturation regions where the drain current is almost constant as well as regions where the drain current is very small, having thus on/off ratio of up to $10^8$ at room temperature. In addition, the nanoperforated GFETs have orders of magnitude higher photoresponses than any graphene detectors not incorporating heterostructures with bandgap materials. These results are a consequence of the fact that e-beam lithography assures pattern uniformity over large areas.

## 2. Fabrication of nanopatterned GFETs

The CVD graphene layer used for the fabrication of GFETs was produced by GRAPHENEA. Graphene was deposited on a (100) silicon substrate of $p$ type, with a resistivity of 8-10 $\Omega \cdot$cm, covered with a 300-nm-thick layer of thermally grown silicon dioxide. The fabrication technology of encapsulated GFETs involved several steps, as follows, the figures in the parentheses illustrating the state of the device after the respective fabrication step:

(1) patterning of the alignment marks, (2) e-beam evaporation of Cr (10 nm) and Au (100 nm), (3) lift-off, (4) patterning of the graphene layer (Fig. 1(a)), including patterning of the hole array (shown in Fig. 1(b)), (5) reactive ion etching of graphene, (6) patterning of source and drain contacts (Fig. 1(c)), (7) e-beam evaporation of Ti (10 nm) and Au (100 nm), (8) lift-off, (9) patterning of the gate insulating layer (Fig. 1(d)), (10) patterning of the gate contact, (11) e-beam evaporation of Ti (10 nm) and Au (100 nm), (12) lift-off (Fig. 1(e)), (13) substrate dicing and encapsulation, and wire bonding (Figs. 1(f))

The detailed SEM image of the nanopatterned graphene channel, fabricated in steps (4) and (5) and displayed in Fig. 1(b), shows a regular array, with an average hole diameter of 125 nm and a neck size of 75 nm. The patterning was performed by e-beam lithography, using a dedicated tool: RAITH e-line. A positive electronresist (PMMA) was used in all patterning steps except the patterning of the gate insulating layer, in which a negative electronresist (HSQ) was utilized. The exposure was done at 30 kV accelerating voltage in all situations. The clearing dose was 300 $\mu C/cm^2$ for the exposed area and 0.8 pC/dot for the perforated holes array in graphene. A highly directional e-beam evaporation equipment for metal deposition: TEMESCAL FC-2000 assured a good quality of the lift-off process. The alignment marks fabricated in the first steps were used for each patterning process in order to achieve good overlay accuracy.

## 3. Electrical measurements of GFETs and discussions

We have performed measurements of all encapsulated transistors using Keithley 4200 SCS equipment with low noise amplifiers at outputs. The DC probes connected to the Keithley 4200 are positioned in a Faraday cage, together with the entire probe station. The graphene FETs are placed on a support located on the chuck of the probe station, which connects the GFETs metallic terminals with DC probes, and then the Faraday cage is closed. The source and the transistor capsule are grounded. The measurements are assisted by a compute, which monitors the entire system. All measurements are performed at room temperature. The validity of our measurements is further tested by repeating the experiments at different voltage steps. No smoothing procedures are used during the measurements or afterwards. The only difference between the measured devices consists in more or less pronounced saturation regions, but otherwise the on/off ratio of all GFETs was greater than $10^6$. Therefore, we present in the following measurements results of two GFETs, considered typical for all devices.

In Fig. 2, we have represented the drain current-drain voltage, $I_D - V_D$, characteristics of a nanoperforated GFET with less pronounced saturation regions at various gate voltages $V_G$ indicated in the inset. From Fig. 2 it follows that the drain current tends to saturate at some gate voltages, e.g. at 1 V and -1 V, a clear negative differential resistance region appears at $V_G = $ -3 V and -4 V, and the transistor is off at the gate voltages of +5 V (current too small to be seen on Fig. 2) where the on/off ratio is $10^8$.

Further, we calculated the drain conductance $g_D$ and the transconductance $g_m$ of the nanoperforated GFET, and have represented $g_D$ as a function of $V_D$ at different gate voltages, and $g_m$ as a function of $V_G$ at various drain voltages in Figs. 3(a) and 3(b), respectively. Unlike in the majority of GFETs, which do not amplify too much even at low frequencies because their amplification, given by the ratio $g_m / g_D$, is small, even less than 1, in our GFETs the transconductance can be substantially higher than the drain conductance over wide ranges of bias voltages. For example, $g_D$ is less than 0.07 mS at all drain voltages for $V_G = 4$V, whereas the modulus of $g_m$ is at least twice as large at the same gate voltage for $V_D$ values starting with 0.5 V, at attains the value of about 0.8 mS for $V_D = 3$ V, suggesting the possibility of high amplification over a wide range of drain voltages. Again, $g_D$ is lower than 0.23 mS at $V_G = 1$ V in the whole range of drain voltages in Fig. 3(a), while the modulus of the corresponding $g_m$ at $V_G = 1$ V is larger than this value for drain voltages above 1.25 V, suggesting the possibility of high amplification over a wide range of $V_D$ values.

In another GFET, the $I_D - V_D$ dependence of which is shown in Fig. 4 at various gate voltages indicated in the inset, more pronounced saturation regions and off regions are clearly visible and the on/off ratio is about $10^7$. The data are represented in this case only for negative gate voltages to evidence clearly the saturation and off regions. The channel is closed at the gate voltage of -5 V for drain voltages not exceeding +4 V.

For this nanoperforated GFET, the $g_D - V_D$ characteristics at various gate voltages indicated in the inset, and the $g_m - V_G$ characteristics at different $V_D$ values given in the inset are displayed in Figs. 5(a) and 5(b), respectively. From Fig. 5(a) it follows that $|g_D| < 0.2$ mS when $V_D > 3$ V for all gate voltages smaller or equal to -2 V, while from Fig. 5(b) it can be seen that $g_m$ is higher than 0.75 mS in these voltage ranges, and attains even 1.07 mS at $V_G = $ -3 V for drain voltages higher than 3 V. Therefore, high amplification is expected to occur also for this GFET over wide ranges of bias voltages.

The high on/off ratio of the GFETs reported in this paper is due to the formation of an energy bandgap in the nanoperforated graphene channel. While this result is well known (see, for example, Refs. [9,10]), it can only have full impact on GFETs with nanoperforated graphene channels if the uniformity of the hole array can be maintained over large areas. As can be seen from Refs. [1,11], where nanoperforated graphene was used as GFET channel, the large-scale uniformity is not preserved, the structure resembling/behaving more as a polycrystalline (or even amorphous) material with holes playing the role of atoms rather than a crystalline material with ordered holes as in our case (see Fig. 1(b)). As a result, high on/off ratio can only be obtained in ordered nanopatterened graphene, as demonstrated by our measurement results. The appearance of a bandgap in the nanoperforated GFET explains also the negative differential resistance regions in Figs. 2 and 4, the transmission coefficient of electrons being minimum (and, hence, the current showing a valley) when the range of energies of the incident electrons has a maximum overlap with the induced bandgap. In graphene, the energy of electrons can be tuned via a gate voltage, and so the $I_D - V_D$ characteristics depend strongly on $V_G$ even in the non-ballistic transport regime.

## 4. Photoresponse of nanoperforated GFETs

The results of the previous section demonstrated that nanopatterning of hole arrays improves the electrical characteristics of GFETs. Because graphene is used also for photodetection, we investigated the photoresponse of the nanoperforated GFET. We used a red laser diode with a power of $P = 0.5$ mW as light source, and have measured the drain current at dark and under illumination. The corresponding $I_D - V_D$ characteristics are shown in the inset of Fig. 6, which illustrates the dependence of the relative photoresponse $\Delta I / I_{D,dark} = I_D - I_{D,dark}$ on the drain voltage. This parameter has impressive values, being as high as 0.35. Note the typical ambipolar, although not symmetric (due to the asymmetric bandgap location), $I_D - V_D$ characteristic of the GFET. The responsivity of the device, defined as $R = \Delta I / P_{eff}$, where $P_{eff} = (A_{graph} / A_{laser})P$, with $A_{graph}$ and $A_{laser}$ the areas of graphene and of the laser beam, has been found to be $R = 10^3$ A/W for $\Delta I = 40$ μA, at $V_D = 1$ V. The area of the laser beam was determined knowing that the diameter of the laser diode is about 1.5 mm, and $A_{graph}$ was estimated from the area of the nanoperforated channel, i.e. 20×5 μm$^2$, taking into account that the holes occupy about 30% of the area. This value of the responsivity is orders of magnitude higher than for any other room-temperature photodetector based only on graphene [12] and on graphene inks [13], for which $R$ << 1 A/W, and is also much larger than in nanostructured graphene photodetectors consisting of quantum-dot like arrays and defect midgap states, in which $R = 8.61$ A/W [14]. The high responsivity value in our nanoperforated GFET can be attributed to the effective electrical field generated by the localized edge states [15]. This built-in electric field induces also a $p$-type doping of the nanoperforated graphene channel, which can be observed from the electrical characterizations in the previous section.

## 5. Conclusions

We have demonstrated that ordered arrays of nanoholes in graphene can be patterned on a wafer scale by e-beam lithography. Nanopatterned graphene, when used as channel in a GFET, improves with orders of magnitude the on/off ratio of the device due to the opening of a bandgap in the nanoperforated channel. In addition, the built-in electric field due to localized edge states at the holes increases with orders of magnitude the photoresponse of nanopatterned GFETs. Thus, GFETs with nanoperforated channels, which can be fabricated at the wafer scale, become competitive devices in nanoelectronics and optoelectronics.

*Acknowledgements* Adrian Dinescu would like to thank prof. Stefano Bellucci from INFN-LNF Frascatti, Italy for discussions about the subject of this paper. We acknowledge the financial support of Nucleus Project TEHNOSPEC PN 1632 2015-2017.

**Figure captions**

Fig. 1 Optical images after different fabrication steps of GFETs with patterned graphene channels with a length of 20 μm and a 5-μm width: (a) PMMA patterning of the graphene, (b) showing a detailed SEM image of the nanopatterned channel with 115 nm average hole diameter of and 70 nm neck size, (c) source and drain contacts as patterned in PMMA, (d) gate insulating layer after patterning with 50-nm-thick HSQ, (e) gate contact after metal deposition and lift-off, and (f) the encapsulated GFET.

Fig. 2 $I_D - V_D$ characteristics of a nanoperforated GFET at various gate voltages indicated in the inset.

Fig. 3 (a) $g_D - V_D$ characteristics of the nanoperforated GFET in Fig. 3 at various gate voltages indicated in the inset, and (b) $g_m - V_G$ characteristics of the same GFET at different $V_D$ values given in the inset.

Fig. 4 $I_D - V_D$ dependence of a nanoperforated GFET at various gate voltages indicated in the inset.

Fig. 5 (a) $g_D - V_D$ characteristics of the nanoperforated GFET in Fig. 5 at various gate voltages indicated in the inset, and (b) $g_m - V_G$ characteristics of the same GFET at different $V_D$ values given in the inset.

Fig. 6 Relative photoresponse of a nanoperforated GFET. Inset: $I_D - V_D$ characteristics at dark and under illumination.

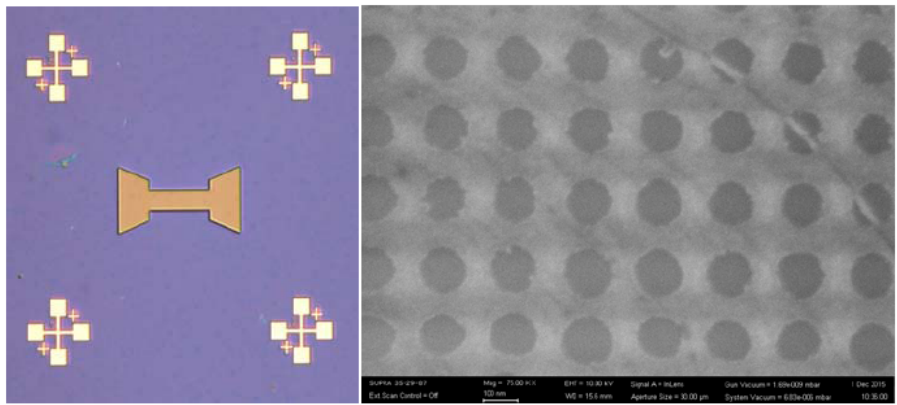

(a)

(b)

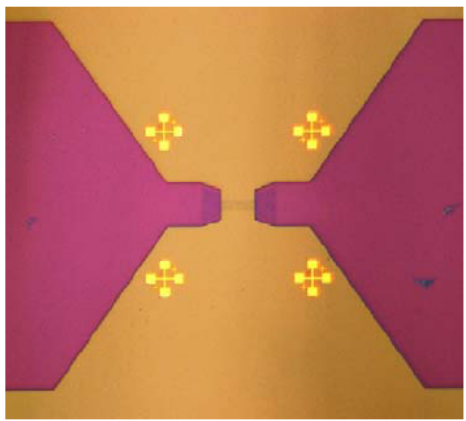

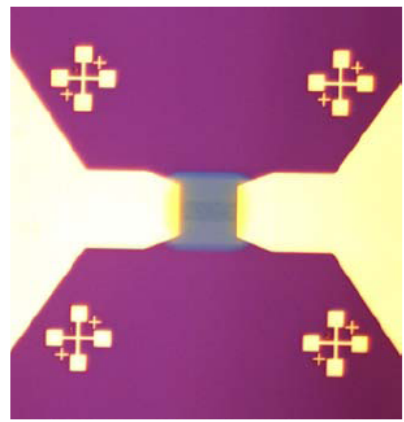

(c)

(d)

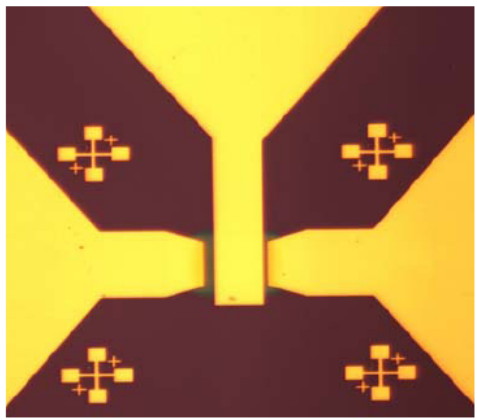

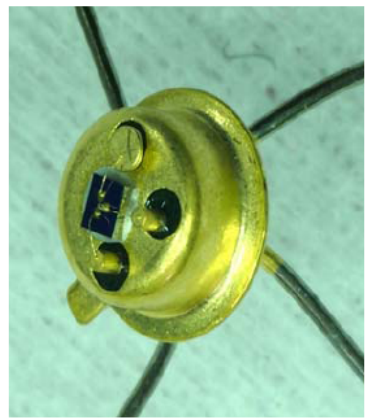

(e)

(f)

Fig. 1

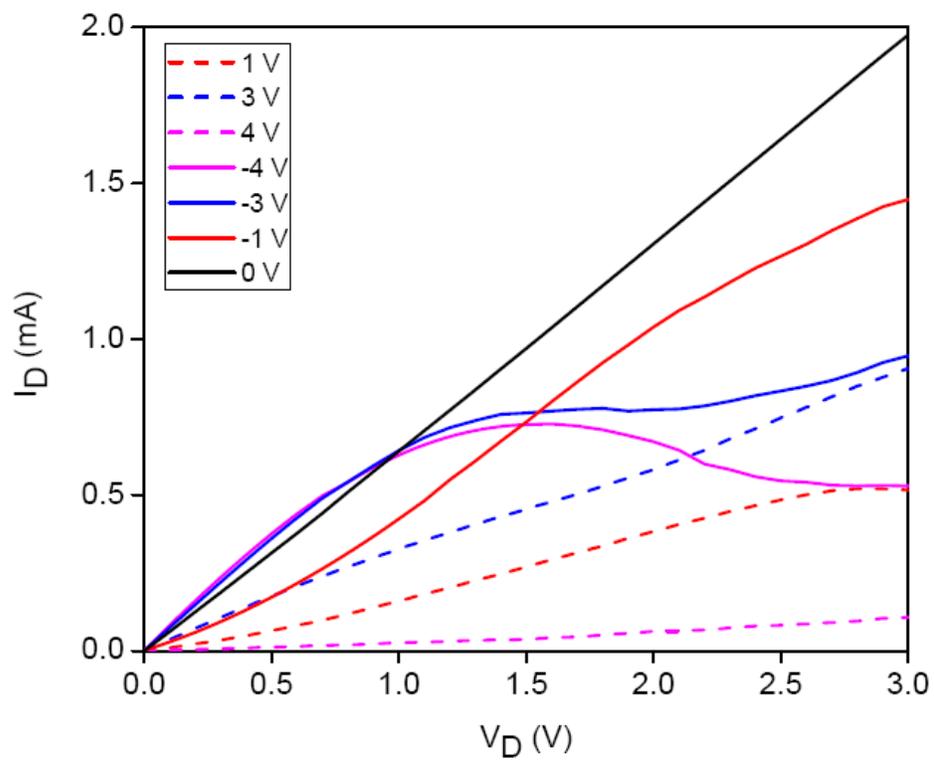

Fig. 2

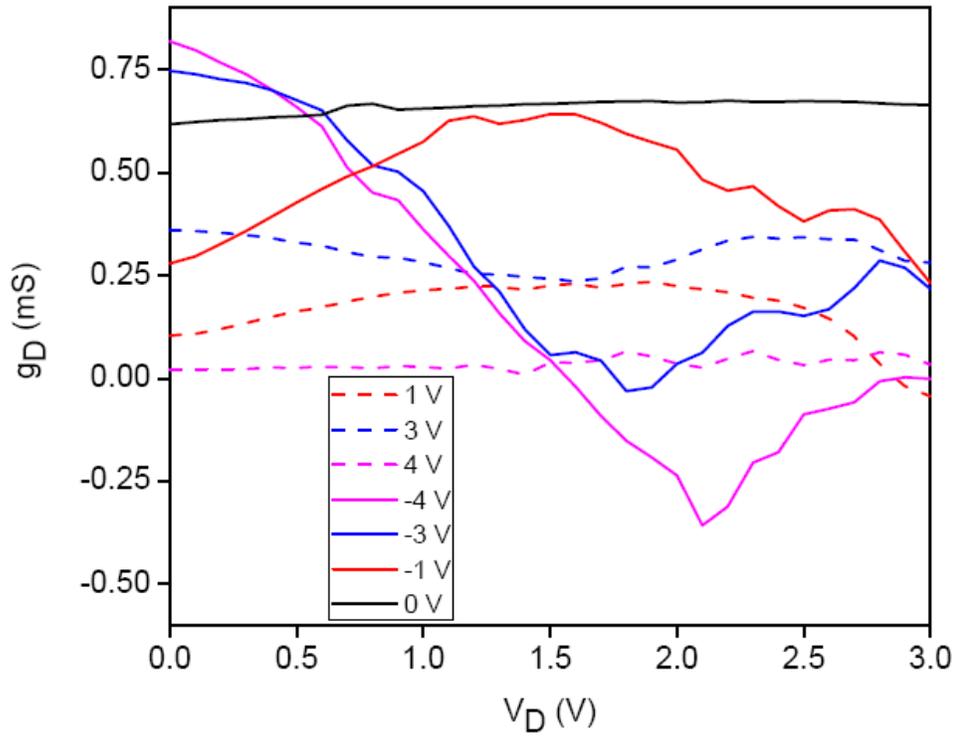

(a)

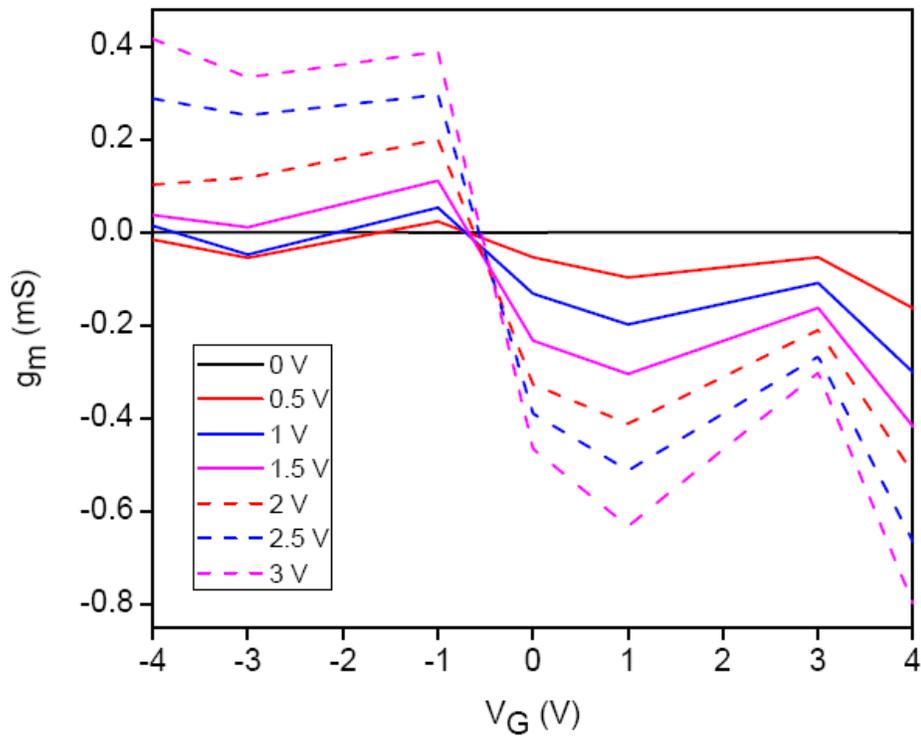

(b)

Fig. 3

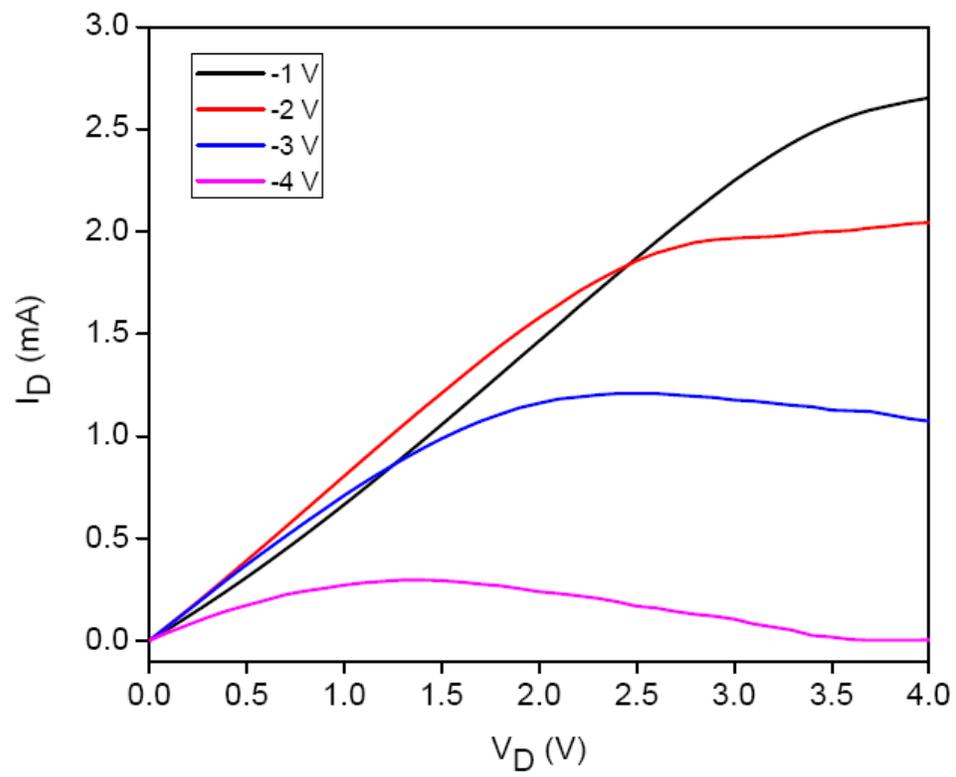

Fig. 4

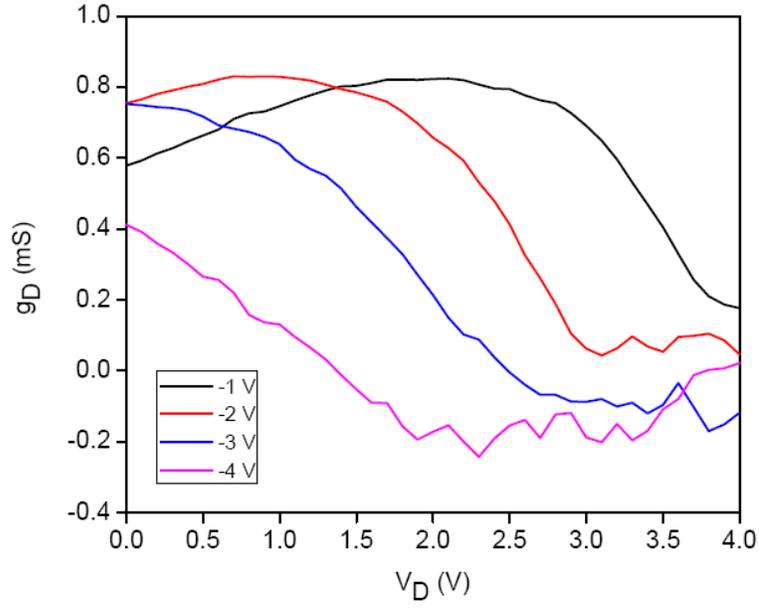

(a)

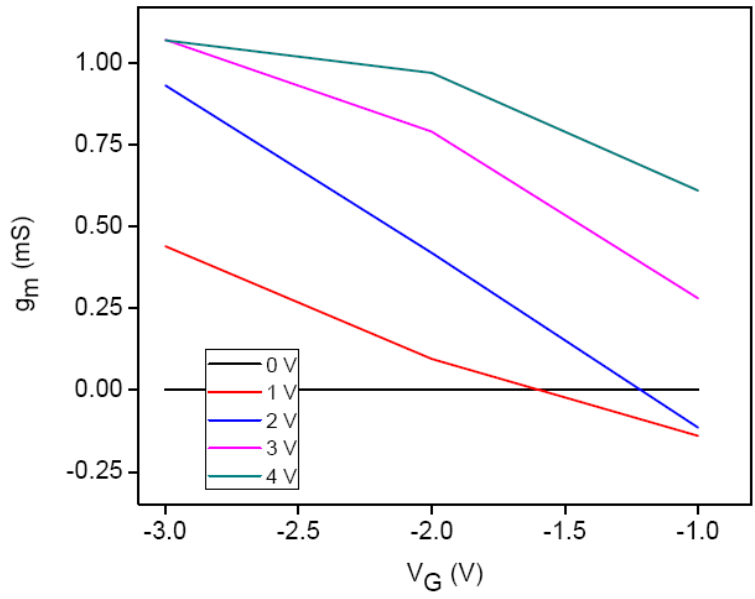

(b)

Fig. 5

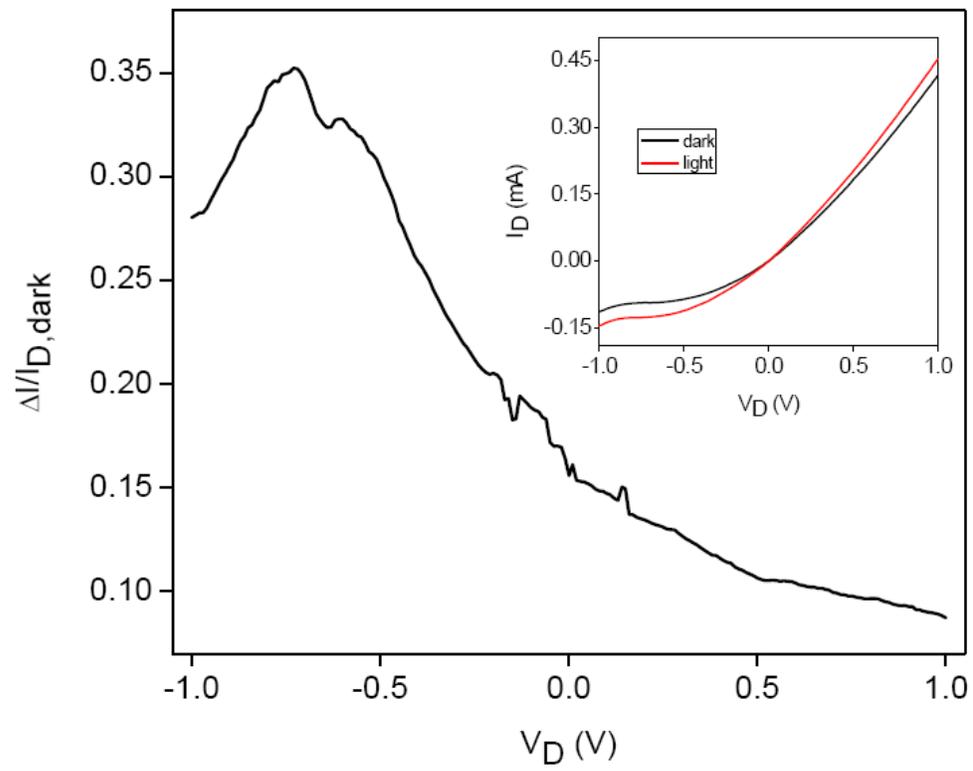

Fig. 6